\documentstyle[12pt]{article}

\begin{document}

\begin{center}
{\large \bf Radius-Mass Scaling Laws for Celestial Bodies}
\end{center}

\begin{center}
{\large R. Muradian, S. Carneiro and R. Marques}
\end{center}

\begin{center}
{\it Instituto de F\'{\i}sica, Universidade Federal da Bahia\\
40210-340, Salvador, BA, Brasil}
\end{center}

\begin{abstract}
In this letter we establish a connection between two-exponent
radius-mass power laws for cosmic objects and previously proposed
two-exponent Regge-like spin-mass relations. A new, simplest method for
establishing the coordinates of Chandrasekhar and Eddington points is
proposed.
\end{abstract}

In previous papers \cite{RM1},\cite{RM2} Muradian has suggested the
two-exponent Regge-like relation

\begin{equation}
J=\hbar \left( \frac m{m_p}\right) ^{1+1/n}  \label{J}
\end{equation}
between the observed mass $m$ and angular momentum $J$ of celestial
bodies.
In this relation $\,\hbar \,$and\thinspace $m_p$ stand, respectively,
for
the Planck constant and for the proton mass. The exponent $n=3$ for
star-like objects and $n=2$ for multistellar ones, like galaxies and
clusters of galaxies.

Relation (\ref{J}), besides to fit reasonably well the observational
data
(see Figure 1), presents two remarkable points: equating (\ref{J}) to
Kerr
limit $J^{Kerr}=Gm^2/c$ for the angular momentum of a rotating black
hole,
we obtain

\begin{equation}
m=m_p\left( \frac{\hbar c}{Gm_p^2}\right) ^{\frac n{n-1}}  \label{m}
\end{equation}
which, for $n=3,$ can be identified with Chandrasekhar mass $%
m_{Ch}=m_p\left( \frac{\hbar c}{Gm_p^2}\right) ^{3/2}$ and, for $n=2$,
with
the Eddington mass $m_E=m_p\left( \frac{\hbar c}{Gm_p^2}\right) ^2$. The

corresponding limiting angular momenta could be obtained by substitution
of
these expressions into (\ref{J}) (or into Kerr relation), as it has been

shown in \cite{RM1}, \cite{RM2}: $J_{Ch}=\hbar \left( \frac{\hbar
c}{Gm_p^2}%
\right) ^2$ and $J_E=\hbar \left( \frac{\hbar c}{Gm_p^2}\right) ^3$.

In a recent paper \cite{Perez}, P\'erez-Mercader has suggested the
existence
of two-exponent scaling relation between mass and radius of cosmic
objects.
Now we will try to establish a connection between such relation and the
above referred Regge-like trajectories in the $J-m$ plane.

First of all, let us note that a theoretical relation between $m$ and
$r$
should be valid, in particular, for Chandrasekhar and Eddington points,
for
which the following relation is valid \cite{SC}

\begin{equation}
r=r_p\left( \frac m{m_p}\right) ^{1/n}=\frac \hbar {m_pc}\left( \frac
m{m_p}\right) ^{1/n}  \label{r}
\end{equation}
where $r_p=\hbar /m_pc$ stands for the proton radius.

Here, for $n=3$ and $m=m_{Ch}$ we obtain the radius of neutron star $%
r_{NS}=\frac \hbar {m_pc}\left( \frac{\hbar c}{Gm_p^2}\right) ^{1/2}$,
while
for $n=2$ and $m=m_{E}$ the radius of observable Universe follows, $%
r_U=\frac \hbar {m_pc}\frac{\hbar c}{Gm_p^2}.$ In this last case
relation (%
\ref{r}) is just a possible expression for well known large number
coincidences \cite{SC}, \cite{FR}.

In this way, (\ref{r}) seems to be a good candidate for theoretical
two-exponent law relating the radius and mass of primordial dense
proto-objects, from which the present day cosmic bodies originated, in
the
sense of Ambartsumian cosmogony (see \cite{RM1} and references there
in).
Another observation in favor of this suggestion is connected to the fact

that the same relations for $r_{NS}$ and $r_U$ follow from the
expression
for half of the gravitational Schwarzchild radius $r=Gm/c^2$ after
substitution of Chandrasekhar $m_{Ch}$ or Eddington $m_E$ masses. This
is
consistent with the above mentioned fact that the Chandrasekhar and
Eddington points correspond in the $J-m$ Chew-Frautschi plane to
maximally
rotating black holes (see Figure 1).

Relation (\ref{r}) is plotted in Figure 2 together with the
observational data
and line $r=Gm/c^2.$ As expected, neutron star and Universe lie on this
black hole line. The theoretical line $r=\frac \hbar {m_pc}\left( \frac
m{m_p}\right) ^{1/2}$ fits crudely the data relating to clusters of
galaxies, but \thinspace in the case of star-like objects the
theoretical $%
\, $relation$\,\,r=\frac \hbar {m_pc}\left( \frac m{m_p}\right) ^{1/3}$
is
completely uncorrelated with the data (except the point of neutron
star).
The following reasoning can elucidate this disagreement.

To (\ref{J}) and (\ref{r}) be consistent, one needs $J=mcr,$ what means
that
(\ref{r}) refers to maximally rotating objects. As we have seen, this is
the
case for the Chandrasekhar and Eddington points, for which the equation
$%
mcr=Gm^2/c$ is equivalent to $r=Gm/c^2$. But, in general, celestial
bodies
are far away from this limit and, in consequence, their radii are
systematically distributed above the lines representing (\ref{r}) (see
Figure 2).

But if (\ref{r}) does not exactly represent the observational data, what

does it represent? And why its $J-m$ partner, relation (\ref{J}), fits
well
the data? A possible answer to these questions is that relations
(\ref{J})
and (\ref{r}) represent an initial dense stage in the evolution of the
bodies, when they have maximum, Regge-like, angular momenta for some
given
radii. So, as bodies evolve, their radii change, diverging from the
original
values given by (\ref{r}).

As indicated in \cite{Perez}, the fact that there are two radically
different power laws for two classes of objects could serve as
indication
that the objects within each class have a similar physical origin. A
possible reason for exponents change is different geometrical shape of
the
primordial objects: disk-like $(n=2)$ for multistellar objects and
ball-like
$(n=3)$ for stellar ones \cite{RM1},\cite{RM2},\cite{FR}.

\end{document}